\documentstyle[eqsecnum,epsfig,twocolumn,aps,floats,graphicx]{revtex}

\begin{document}
\draft \preprint{}

\twocolumn[\hsize\textwidth\columnwidth\hsize\csname
 @twocolumnfalse\endcsname

\title{Electronic structure of the trilayer cuprate superconductor
 Bi$_2$Sr$_2$Ca$_2$Cu$_3$O$_{10+\delta}$}
\author{D.L. Feng$^{1}$, A. Damascelli$^{1}$, K.M. Shen$^{1}$, N.
Motoyama$^2$,  D.H. Lu$^1$, H. Eisaki$^1$, K. Shimizu$^3$, J.-i.
Shimoyama$^3$, K. Kishio$^3$, N. Kaneko$^1$, M. Greven$^{1}$, G.D.
Gu$^4$, X.J. Zhou$^1$, C. Kim$^1$, F. Ronning$^{1}$, N.P.
Armitage$^{1}$, Z.-X. Shen$^{1}$}
\address{$^1$Department of Physics, Applied Physics, and Stanford Synchrotron Radiation
Laboratory,\\ Stanford University, Stanford, California 94305}
\address{$^2$Department of Superconductivity, University of Tokyo, Tokyo,
113-8656, Japan}
\address{$^3$Department of Applied Chemistry, University of Tokyo, Tokyo,
113-8656, Japan}
\address{$^4$Physics Department, Brookhaven National Laboratory, P. O.
Box 5000, Upton, New York 11973}

\date{\today}
\maketitle

\begin{abstract}
The low-energy electronic structure of the trilayer cuprate
superconductor Bi$_2$Sr$_2$Ca$_2$Cu$_3$O$_{10+\delta}$ near
optimal doping is investigated by angle-resolved photoemission
spectroscopy. The normal state quasiparticle dispersion and Fermi
surface, and the superconducting $d$-wave gap and coherence peak
are observed and compared with those of single and bilayer
systems. We find that both the superconducting gap magnitude and
the relative coherence-peak intensity scale linearly with $T_c$
for various optimally doped materials. This suggests that the
higher $T_c$ of the trilayer system should be attributed to
parameters that simultaneously enhance phase stiffness and pairing
strength.
\end{abstract}

 \vspace{-.55cm}

\vskip2pc
 ]
  \narrowtext

The high-$T_c$ cuprate superconductors (HTSCs), based on the
number of CuO$_2$ planes in the characteristic multilayer blocks,
can be classified into single-layer materials [e.g.,
Bi$_2$Sr$_2$CuO$_{6+\delta}$ (Bi2201), HgBa$_2$CuO$_{4+\delta}$
(Hg1201), and La$_{2-x}$Sr$_x$CuO$_{4}$ (LSCO)], bilayer materials
[e.g., Bi$_2$Sr$_2$CaCu$_2$O$_{8+\delta}$ (Bi2212),
HgBa$_2$CaCu$_2$O$_{6+\delta}$ (Hg1212) and
YBa$_2$Cu$_3$O$_{7-\delta}$ (Y123)], trilayer materials [e.g.,
Bi$_2$Sr$_2$Ca$_2$Cu$_3$O$_{10+\delta}$ (Bi2223), and
HgBa$_2$Ca$_2$Cu$_3$O$_{8+\delta}$ (Hg1223)], and so on. This
structural characteristic has a direct correlation with the
superconducting properties: within each family of cuprates, the
superconducting phase transition temperature ($T_c$) increases
with the layer number ($n$) for $n\!\leq\!3$, and then starts to
decrease\cite{Stasio90,Tarascon92}. Taking the Bi-family of HTSCs
as an example, the maximum $T_c$ is approximately 34, 90, and
110\,K for optimally doped Bi2201 ($n\!=\!1$), Bi2212 ($n\!=\!2$),
and Bi2223 ($n\!=\!3$), respectively. Despite various experimental
and theoretical efforts, a conclusive microscopic understanding of
this evolution has not yet been reached, partly because of the
lack of detailed knowledge about the electronic structure of the
trilayer systems. In particular, angle-resolved photoemission
spectroscopy (ARPES), one of the most direct probe of the
electronic structure of HTSCs \cite{ARPESHTSC}, has so far been
limited to single and bilayer compounds. To gain further insight
into the role of multiple CuO$_2$ planes in determining the
macroscopic physical properties of the cuprates, like the value of
the $T_c$, it is crucial to extend the investigation of the
electronic structure to trilayer HTSCs, and to compare the results
with those from the single and bilayer materials. Given that the
Bi-based cuprates represent the HTSC family best characterized by
ARPES, the trilayer system Bi2223 is the ideal candidate for such
a comparative study.

In this Letter, we report the first ARPES study, to the best of
our knowledge, of the electronic structure of the trilayer HTSC
Bi2223, for which high quality single crystals with dimensions
suitable for ARPES measurements has been recently synthesized. As
in the single- and bi-layer materials, at nearly optimally doped
Bi2223, we observed a large hole-like Fermi surface, a flat
quasiparticle band near $(\pi,0)$, $d$-wave pseudo and
superconducting gaps, and a large superconducting peak (the
so-called coherence peak in the case of Bi2212). The
superconducting gap magnitude and the relative weight of the
superconducting peak both increase linearly with $T_c$ for the
optimally doped Bi-based HTSCs. This indicates that the higher
$T_c$  of Bi2223 is caused by the enhancement of both pairing
strength and phase stiffness, consistent with the idea that
optimal doping corresponds to the intersection between
phase-coherence and pairing-strength temperature scales.

Bi2223 single crystals were grown by floating-zone technique.
Nearly optimally doped samples [$T_c\!=\!108$\,K, $\Delta
T_c(10\%\!-\!90\%)\!=\!2$\,K] were obtained by subsequently
annealing the slightly underdoped as-grown Bi2223 crystals
($T_c=105$\,K) for three days at 400\,$^\circ$C and
$P_{O_2}\!=\!2.1$\,atm, and then rapidly quenching them to room
temperature. Magnetic susceptibility measurements did not detect
the presence of second phases, and X-ray diffraction showed well
ordered bulk structures, with the typical superstructure seen in
Bi2201 and Bi2212. Optimally doped Bi2212 ($T_c=90$\,K) and Bi2201
($T_c=34$\,K) with $\Delta T_c(10\%\!-\!90\%)\!=\!1$\,K were also
studied for comparison. ARPES experiments were performed at the
Stanford Synchrotron Radiation Laboratory (SSRL) on a beamline
equipped with a Scienta SES200 electron analyzer. Multiple ARPES
spectra were acquired simultaneously in a narrow window of
$0.5^\circ\!\times\!14^\circ$ with, unless otherwise specified, an
angular resolution of $0.3^\circ$ (along the cut direction) and an
energy resolution of 10\,meV. The samples were aligned by Laue
diffraction, and cleaved {\it in-situ} under a pressure better
than $5\!\times\!10^{-11}$ torr. Bi2223 samples \#1,\,\#3
(\#2,\,\#4) were cleaved at $T\!=\!10$\,K ($T\!=\!125$\,K). Data
were collected within 12 hours after cleaving and aging effects
were negligible.

\begin{figure}[t!]
\centerline{\epsfig{file=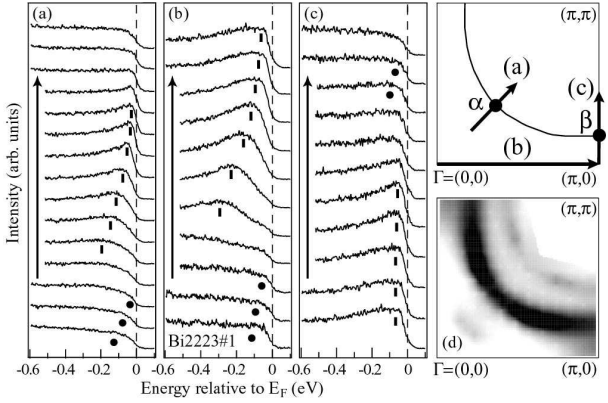,clip=,angle=0,width=8cm}} \vspace{0in}
\caption{(a-c) Normal state Bi2223 ARPES spectra along the high-symmetry lines,
as indicated in the BZ sketch (data taken at 125\,K with 21.2 eV photons and
angular resolution of 0.24$^\circ$, 0.6$^\circ$, and 0.3$^\circ$,
respectively). Main ({\it umklapp}) bands are marked with bars (circles). (d)
Integrated $E_F$-intensity map ($\pm10$ meV) symmetrized with respect to
(0,0)-($\pi$,$\pi$).} \label{dispersion}
\end{figure}

Fig.\,\ref{dispersion} presents the normal state ARPES spectra measured on
Bi2223 along the high symmetry directions of the first Brillouin zone (BZ).
Similar to what has been observed on optimally doped Bi2201 and
Bi2212\cite{ARPESHTSC}, the quasiparticle band is rather flat near ($\pi$,0)
while it is quite dispersive and defines a clear Fermi crossing along the
(0,0)-($\pi$,$\pi$) direction. The {\it umklapp} bands, one of the
characteristics of the Bi-family of cuprates, are also detected. The Fermi
surface (FS) can be identified by the local maxima of the intensity map
obtained by integrating the ARPES spectra within a narrow energy window at the
Fermi energy ($E_F$), after the spectra were normalized with respect to the
high energy spectral weight. As in the case of Bi2201 and
Bi2212\cite{ARPESHTSC}, one main and two weak {\it umklapp} FSs, shifted by
$\pm(0.21\pi,0.21\pi)$ with respect to the main FS, are clearly observed
(Fig.\,\ref{dispersion}d).

\begin{figure}[t!]
\centerline{\epsfig{file=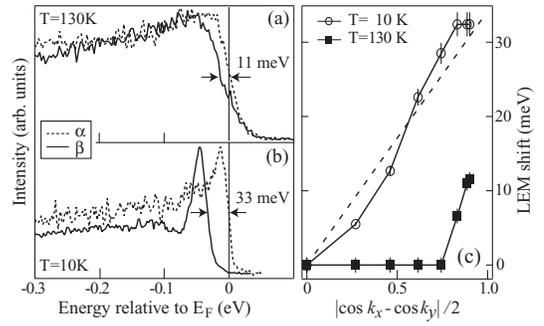,clip=,angle=0,width=7cm}} \vspace{0in}
\caption{(a) Normal and (b) superconducting state Bi2223 spectra measured with
21.2 eV photons at $\alpha$ and $\beta$ (see BZ sketch in
Fig.\,\ref{dispersion}). (c) Position of the leading-edge midpoint (LEM) above
and below $T_c$ along the normal state FS. The dashed line is a fit to the
$d$-wave gap functional form.} \label{gap}
\end{figure}

By tracking the energy position of the leading-edge midpoint (LEM) as a
function of temperature and momentum, one can identify an anisotropic pseudogap
and a superconducting gap ($\Delta$) consistent with a $d$-wave symmetry.
Figs.\,\ref{gap}a and \ref{gap}b show that at $\alpha$, where the FS crossing
along the nodal region is found (see the BZ sketch in Fig.\,\ref{dispersion}),
the LEMs of both normal and superconducting state spectra are located at $E_F$,
indicating the absence of any gap. On the other hand, in the antinodal region
(i.e., at $\beta$) the LEM is always shifted below $E_F$, corresponding to an
11\,meV pseudogap above $T_c$ and a 33\,meV superconducting gap below $T_c$.
The momentum dependence of both normal and superconducting state gaps along the
normal state FS is summarized in Fig.\,\ref{gap}c. The superconducting gap can
be fitted to the $d$-wave functional form $\Delta\!=\!\Delta_{0}|\cos
k_x\!-\!\cos k_y|/2$ (where $\Delta_{0}$ is the superconducting gap amplitude),
while the pseudogap vanishes in wide momentum-space regions resulting in a
partially gapped FS (or, equivalently, four disconnected FS arcs in the BZ) at
130\,K. Similar phenomena have also been observed in Bi2212 \cite{Fermiarc}.
Furthermore, for the Bi2223 samples \#2-4 the pseudogap at $\beta$ was found to
vary from 6 to 9 meV at 125\,K (which is possibly caused by some small
variations in carrier dopings), and the sample with larger pseudogap also has a
larger superconducting gap.

\begin{figure}[t!]
\centerline{\epsfig{file=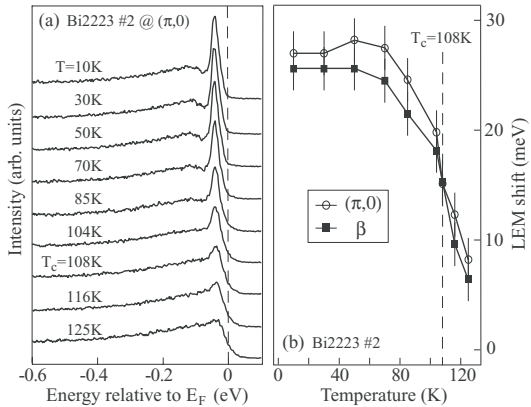,clip=,angle=0,width=7cm}} \vspace{0in}
\caption{(a) Temperature dependence of the Bi2223 $(\pi,0)$ spectra and (b) of
the LEM energy shift at $(\pi,0)$ and $\beta$.} \label{scp}
\end{figure}

Again in analogy with the case of Bi2212\cite{Kaminski00}, in Fig.\,\ref{gap}
one also notices that the normal state spectrum at $\alpha$ sharpens up upon
entering the superconducting state, but the most dramatic change in the
lineshape takes place at $\beta$, where the spectrum evolves into a {\it
peak-dip-hump} structure below $T_c$. This so-called superconducting peak,
which dominates the spectral function in the ($\pi$,0) region, has been argued
to be an important characteristics of the HTSCs\cite{Scp}. It has so far been
detected by ARPES only on Bi2212\cite{Dessau91} and Y123\cite{Lu01}, and the
present results substantiate its existence in the spectral function of an
$n\!=\!3$ system. In order to gain more information, detailed temperature
dependence measurements were performed at $(\pi,0)$, and the results are
presented in Fig.\,\ref{scp}a. The superconducting peak emerges slightly above
$T_c$ (i.e., at 116 K).  Upon further cooling the sample below $T_c$, its
intensity increases rapidly  before it eventually saturates at low
temperatures, while the total spectral weight is conserved (within 1-2\%). At
the same time, the LEM shifts to high binding energies reflecting the opening
of the superconducting gap (Fig.\,\ref{scp}b). Note also that, due to the weak
quasiparticle dispersion in the flat band region, the spectra at $(\pi,0)$ and
$\beta$ exhibit a very similar behavior, as emphasized by Fig.\,\ref{scp}b.

\begin{figure}[b!]
\centerline{\epsfig{file=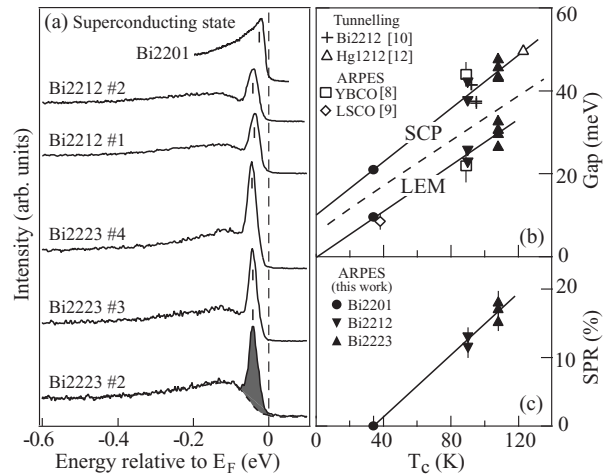,clip=,angle=0,width=7.9cm}} \vspace{0in}
\caption{ (a) Superconducting state $(\pi,0)$ spectra measured at 10\,K  with
22.7 eV photons on optimally doped Bi2201 and Bi2212, and nearly optimally
doped Bi2223. (b) Superconducting gap magnitude as estimated from the position
of the superconducting peak (SCP) and the LEM shift (separated by the dashed
line), for various optimally doped materials, and (c) superconducting peak
ratio (SPR) extracted from the data in (a), plotted versus $T_c$.}
\label{compare}
\end{figure}

So far, we have shown that various properties of Bi2223 qualitatively resemble
those of Bi2212 and/or Bi2201. The natural question is: what part of the
electronic structure of Bi2223 can account for the highest $T_c$ among the
Bi-family of cuprates? To further investigate this issue, we compare in
Fig.\,\ref{compare}a the superconducting state $(\pi,0)$ spectra from optimally
doped Bi2201 and Bi2212, and nearly optimally doped Bi2223 taken under the same
experimental conditions (except for the higher energy resolution, i.e. 6meV,
used for the Bi2201 data). The superconducting gap magnitude  $\Delta_{0}$ can
be estimated by either the position of the superconducting peak or the LEM
shift below $E_F$ in the ($\pi$,0) spectra. We found that the average LEM (peak
position) gap values are 10 (21), 24 (40), 30 (45) meV for the $n\!=\!1,2,3$
systems, respectively. As shown in Fig.\,\ref{compare}b,  the gap value of the
three different systems scales linearly with the corresponding $T_c$. In
particular, the LEM gap can be well fitted by a line across the origin
corresponding to an $n$-independent ratio $2\Delta_0/k_BT_c\!\simeq\!5.5$.
Furthermore, from ARPES and tunnelling spectroscopy results reported for other
families of cuprates it is found that the values of $\Delta_{0}$ for optimally
doped LSCO\cite{LSCOgap}, Bi2212\cite{Bi2212tunneling}, YBCO\cite{Lu01,chains},
and Hg1212\cite{Hggap} follow the same gap versus $T_c$ linear relation (see
Fig.\,\ref{compare}b).

From the data presented in Fig.\,\ref{compare}a, one can also
extract the so-called superconducting peak ratio (SPR), which is
defined as the ratio between the integrated spectral weight of the
superconducting peak and that of the whole spectrum (i.e.,
from\,$-$0.5 to\,+0.1\,eV). As shown in Fig.\,\ref{compare}a, for
the Bi2223 sample \#2, the peak intensity is obtained by fitting
the smooth ``background" with a phenomenological function and then
subtracting its contribution to the total integrated weight, as
discussed in detail elsewhere\cite{Feng000}. For Bi2201, the
superconducting peak is not resolved in the ARPES data and
therefore its SPR is estimated to be close to zero. In recent
scanning tunnelling spectroscopy (STS) experiments a
superconducting peak in the density of state was observed for
Bi2201. This, however, was detected only at certain locations on
the cleaved sample surface and was not resolved in the spatially
averaged STS spectra\cite{STM}, consistent with what is observed
by ARPES. For Bi2212 and Bi2223, the spectra in
Fig.\,\ref{compare}a (normalized at high binding energy to allow a
direct comparison) indicate that the superconducting peak
amplitude for Bi2223 is much larger than that of Bi2212. Overall,
the SPRs of these systems scale linearly with $T_c$
(Fig.\,\ref{compare}c)\cite{Note}. For Bi2212, it has been argued
that the SPR is related to the phase stiffness of the condensate
or superfluid density ($\rho_s$)\cite{Feng000,Ding000}. The weak
superconducting peak in the (spatially averaged) ARPES spectra
from Bi2201 may then reflect a low superfluid density, and in fact
the peak amplitude is negligible also in Bi2212 samples with
$T_c\!<\!50$\,K\cite{Feng000}. The $n$-dependence of the SPR is
qualitatively consistent with the muon spin resonance ($\mu$SR)
results, which show that $\rho_s$ for the optimally doped cuprates
increases with $n$ (for $n\leq3$), and scales with $T_c$ in
approximately a linear fashion as in the celebrated ``Uemura plot"
\cite{Uemura93}. Therefore, the ARPES results together with those
from tunnelling and $\mu$SR indicate that both $\Delta_{0}$ and
$\rho_s$ increase with $T_c$ for the different optimally doped
cuprates.

\begin{figure}[b!]
\centerline{\epsfig{file=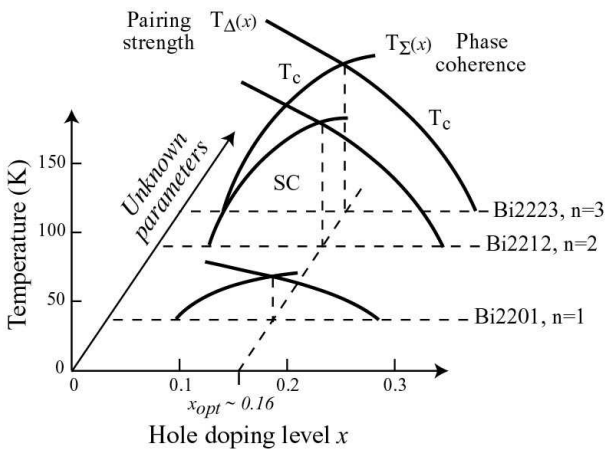,clip=,angle=0,width=8cm}} \vspace{0in}
\caption{Qualitative phase-diagram for the Bi-based HTSCs.} \label{phase}
\end{figure}

Within current understanding, $\Delta_{0}$ and $\rho_s$ are the
two most important quantities in characterizing the
superconducting state, as they reflect the strength of the two
basic ingredients of superconductivity: pairing and phase
coherence. $T_{\Delta}$, the temperature at which the Cooper pairs
start to form, is determined by pairing strength (or $\Delta_0$);
$T_{\Sigma}$, the temperature at which the Cooper pairs, if any,
become phase coherent, is determined by the phase stiffness (or
$\rho_s$). The superconducting phase transition temperature is
given by $T_c\!=\!min(T_{\Delta},T_{\Sigma})$ \cite{phasedaigram}.
For conventional superconductors, $T_{\Sigma}\!\gg\!T_{\Delta}$;
therefore, $T_c\!=\!T_{\Delta}$ and phase fluctuations are not
important in determining $T_c$. The situation is different for the
HTSCs: in order to have high $T_c$, it is necessary to have both
large $\Delta_{0}$ and $\rho_s$, as we have seen for nearly
optimally doped Bi2223. The reason for this is that HTSCs are
doped Mott insulators with low carrier density, for which
$T_{\Sigma}$ and $T_{\Delta}$ are comparable and proposed to have
the doping dependence sketched in Fig.\,\ref{phase}
\cite{phasedaigram,Uemura97}. The crossing of $T_{\Delta}(x)$ ($x$
being doping) and $T_{\Sigma}(x)$ gives $T_{\Delta}(x_{opt})\!=\!
T_{\Sigma}(x_{opt})\!=\!T_{c,opt}$ (with the subscript $opt$
referring to optimal doping, which is found to be approximately
fixed at $x_{opt}\!\simeq\!0.16$ for many HTSCs \cite{doping}).
The approximate linear relations $\Delta_{0,opt}\!\propto\!
T_{c,opt}$ and $\rho_{s,opt} \!\propto\! T_{c,opt}$ observed for
various optimally doped systems lead to $T_{\Sigma}(x_{opt})
\!\propto\! \rho_{s,opt}$ and $T_{\Delta}(x_{opt}) \!\propto\!
\Delta_{opt}$, as theoretically proposed\cite{phasedaigram}.

We have shown that many aspects of the electronic structure of
Bi2223, such as the Fermi surface topology and flat band
dispersion, resemble those of Bi2212 and Bi2201. A preliminary
lineshape analysis\cite{next} suggests that the interlayer
coupling between CuO$_2$ planes within a multilayer block is not
stronger, but possibly even weaker in Bi2223 than in Bi2212, where
bilayer band splitting, which causes multiple features or broader
lineshapes in ARPES spectra, were recently observed\cite{BBS4}.
This and the fact that $T_{c,opt}$ in Hg1201 is comparable to that
of Bi2212 indicate that the interlayer coupling within a
multilayer block is not the dominant factor for the enhancement of
$T_c$. Moreover, $T_{c,opt}$ does not scale with $n$ in a linear
way within a specific HTSC family; and for a given $n$, e.g.
$n=1$, $T_{c,opt}$ varies from $30$\,K to $100$\,K for different
families of cuprates. Instead, we have shown that $T_{c,opt}$
scales approximately linearly with both $\rho_{s,opt}$ and
$\Delta_{0,opt}$. One could speculate that the resolution of the
$T_c$ vs. $n$ problem might be incorporated into a broader task,
namely the search for the parameters that enhance both
superconducting gap and superfluid density, and in turn the
optimal $T_c$. These parameters could be affected by $n$ and other
conspiring factors, for which various candidates have already been
proposed, including superconductivity enhancement in the
non-CuO$_2$ layers\cite{Ted}, or as a consequence of impurities
and distortion/strain introduced into the
system\cite{Hiroshi,Bianconi}. To highlight these unknown
parameters, we add a third axis to the phase diagram of the
hole-doped HTSCs (Fig.\,\ref{phase}), along which both pairing
strength and phase stiffness (and thus $T_{c,opt}$) increase with
the same monotonic trend, contrary to their opposite trends along
the doping axis. In this way, the Bi-based cuprates and possibly
different families of HTSCs can be integrated into one
comprehensive phase diagram.

{\it Acknowledgements:} DLF and ZXS would like to thank S. Maekawa
and T. H. Geballe for helpful discussions. SSRL is operated by the
DOE Office of Basic Energy Science Divisions of Chemical Sciences
and Material Sciences. The Stanford experiments are also supported
by the NSF grant DMR0071897 and ONR grant N00014-98-1-0195-A00002.
The crystal growth work at Stanford was supported by DOE under
Contract Nos. DE-FG03-99ER45773-A001 and DE-AC03-76SF00515. MG is
also supported by the A. P. Sloan Foundation and NSF CAREER Award
No. DMR-9985067.


\begin{references}
\bibitem{chains}For YBCO, because of the additional gap anisotropy due to the presence of the CuO
chains\cite{Lu01}, the maximum gap amplitude (i.e., $\Delta_0$ at
the $Y$ point) is plotted in Fig.4b.
\bibitem{Note}We note that Bi2223 and Bi2212 have different
structures and thus possibly different photoemission matrix
elements. However, the large enhancement of the superconducting
peak in Bi2223 and the qualitative aspect of Fig.\,4c are not
likely just matrix element artifacts.
\bibitem{Uemura97}Y.J. Uemura, Physica C {\bf 282-287},
194 (1997), which also shows that $\rho_s$ (and thus $T_{\Delta}$)
eventually decreases in the overdoped regime. It is omitted here
for simplicity, and will not affect the qualitative conclusions
drawn here.
\end{references}
\end{document}